\documentclass[11pt,a4paper]{article}
\pdfoutput=1
\usepackage{jcappub}

\usepackage[utf8]{inputenc}
\usepackage{float}
\usepackage{amsmath, amssymb}
\usepackage{cancel}

\makeatletter

\usepackage{bm}\usepackage{tensor}
\usepackage{comment}
\usepackage{ifpdf}
\usepackage{slashed}
\usepackage{color}
\usepackage[mathscr]{eucal}

\makeatother

\begin{document}

%%%%%%%%%%%%%%%%%%%%%%%%%%%%%%%%%%%%%%%%%%%%%%%%%%%%%%%%%%%%%
\begin{flushright}
    IPMU19-0134
\end{flushright}
\title{Oscillon of Ultra-Light Axion-like Particle}

\author[a, b]{Masahiro Kawasaki,}
\author[a, b]{Wakutaka Nakano,}
\author[a, b]{and Eisuke Sonomoto}

\affiliation[a]{ICRR, University of Tokyo, Kashiwa, 277-8582, Japan}
\affiliation[b]{Kavli IPMU (WPI), UTIAS, University of Tokyo, Kashiwa, 277-8583, Japan}

\abstract{
Ultra-light axion-like particle (ULAP) is one of attractive candidates for cold dark matter.
Because the de Broglie wavelength of ULAP with mass $\sim 10^{-22} {\rm eV}$ is $\mathcal{O}({\rm kpc})$,
the suppression of the small scale structure by the uncertainty principle can solve the core-cusp problem.
Frequently, ULAP is assumed to be uniformly distributed in the present universe.
In typical ULAP potentials, however, 
strong self-resonance at the beginning of oscillation invokes the large fluctuations, 
which may cause the formation of the dense localized object, oscillon.
% Such a dense object lives for a long time, it may affect the cosmological evolution.
In this paper,
we confirm the oscillon formation in a ULAP potential by numerical simulation
and analytically derive its lifetime.
% and briefly discuss its consequence.
}


\emailAdd{kawasaki@icrr.u-tokyo.ac.jp}
\emailAdd{m156077@icrr.u-tokyo.ac.jp}
\emailAdd{sonomoto@icrr.u-tokyo.ac.jp}
%\notoc

\maketitle
\flushbottom

%%%%%%%%%%%%%%%%%%%%%%%%%%%%%%%%%%%%%%%%%%%%%%%%%%%%%%%%%%%%%
\section{Introduction}
\label{sec:introduction}

Identifying the dark matter is the issue of greatest concern both in particle physics and cosmology. 
Amongst various models, 
cosmological constant and cold dark matter ($\Lambda$CDM) model is the most favored by many observations~\cite{Komatsu:2010fb, Aghanim:2018eyx}.

$\Lambda$CDM model significantly succeeds in explaining large-scale structure of the universe though,
it still has astronomical problems in small scales.
One of the famous problems is the core-cusp problem,
in which the tension between the simulated cusp profile~\cite{Navarro_1997, Fukushige:1996nr, Ishiyama:2011af}
and observed cored profile of the galactic center~\cite{Borriello:2000rv, Gilmore:2007fy, Oh:2008ww} is reported.
This problem is solved, for example,
when dark matter is ultra-light ($m \sim 10^{-22}\ {\rm eV}$)
and coherently oscillating~\cite{Hu:2000ke, Hui:2016ltb}.
This is because the large de Broglie wavelength of the ultra-light particle, approximately the same scale as the galactic center ($\sim {\rm kpc}$), 
smears out the central excessive density.
This particle is often called ultra-light axion-like particle (ULAP)
because it is ultra-light and coherently oscillating like QCD axion~\cite{Weinberg:1977ma, Wilczek:1977pj, Dine:1981rt}.

In this case, 
clumps of ULAP can be formed by the gravity interaction like axion star~\cite{Kolb:1993zz}
(we call it "ULA star" with an analogy to QCD axion).
ULA star is a kind of soliton 
that the attractive forces from gravity and repulsive forces from the kinetic pressure are balanced.
Some studies show 
that ULAP forms ULA star by gravity interaction at the galactic center~\cite{Schive:2014dra, Schive:2014hza, Veltmaat:2018dfz}.
The existence of the star solution is also proved in the contest of axion~\cite{Braaten:2015eeu, Visinelli:2017ooc}
(the axion star formation is also confirmed in Ref.~\cite{Levkov:2018kau}).

Before the ULA star formation, however,
self-interaction of ULAP can lead to the oscillon~\cite{Bogolyubsky:1976yu, Gleiser:1993pt, Copeland:1995fq} formation by parametric resonance~\cite{Shtanov:1994ce, Kofman:1995fi, Kofman:1997yn,Yoshimura:1995gc} in the radiation dominated universe.
Oscillon is a spatially localized solitonic state of a real scalar field,
which often survives for a long time because of the conservation of the adiabatic invariance~\cite{Kasuya:2002zs,Kawasaki:2015vga}.
The existence of such dense objects in the universe
affects the current observational constraint~\cite{Li:2013nal, Nori:2018pka, Porayko:2018sfa} 
and proposed experiments~\cite{Graham:2017ivz, Nagano:2019rbw}.
% The oscillon formation affects the  structure formation
% ULAP star formation (oscillon seed)
% dark matter detection.
% new detection method

The possibility of long lifetime oscillons in a ULAP potential is pointed out in Ref.~\cite{Olle:2019kbo},
but their formation and precise lifetime is unclear
because analytical estimation of the fluctuation growth and lifetime is difficult.
In this paper,
we confirm the oscillon formation by classical lattice simulation in a particular parameter region.
We also calculate the oscillon lifetime by the analytical method derived in Ref.~\cite{Ibe:2019vyo},
which shows that produced ULAP oscillons live long, maybe up to the present day.

In Sec.~\ref{sec:oscillon}, 
we show that the condition for the oscillon formation is satisfied in a typical ULAP potential.
In Sec.~\ref{sec:formation},
we numerically confirm the oscillon formation by the classical lattice simulation.
In Sec.~\ref{sec:lifetime},
we analytically derive the lifetime of produced oscillon and briefly discuss the cosmological consequence.
Finally, in Sec.~\ref{sec:conclusion},
we conclude our results.

%%%%%%%%%%%%%%%%%%%%%%%%%%%%%%%%%%%%%%%%%%%%%%%%%%%%%%%%%%%%%
\section{Oscillon of Ultra-Light Axion Particle}
\label{sec:oscillon}

In the context of ULAP, we often use the cosine potential with an analogy to axion. 
In that case, however, large initial fluctuations are necessary for the oscillon formation
because the instability generated by the cosine potential is too weak to enhance the fluctuations.
In addition,
the resultant oscillon is unstable
(the lifetime is $\mathcal{O}(100) m$)~\cite{Piette:1997hf}
and hardly leave interesting cosmological effects~\cite{Vaquero:2018tib, Buschmann:2019icd}.

Still many other potentials of ULAP are suggested~\cite{Dong:2010in, Kallosh:2013hoa, Kallosh:2013tua, Kallosh:2013yoa},
instead of the cosine potential we use the following potential~\cite{Silverstein:2008sg, McAllister:2008hb, Nomura:2017ehb}
\begin{equation}
    V(\phi) = \frac{m^2F^2}{2p} \left[1 - \left(1+\frac{\phi^2}{F^2}\right)^{-p} \right],
    \hspace{5mm} (p > -1)
    \label{eq:potential}
\end{equation}
where $p > -1$ is necessary to guarantee the existence of the oscillon solution
(see \ref{sec:solution} and Eq.~(\ref{eq:oscillon-condition})).
In the following subsections,
we will show that the two general conditions necessary for the oscillon formation are satisfied;
% we will show that the two general conditions for the oscillon formation
%%
\begin{enumerate}
   \item The existence of the oscillon solution,
   \item The large fluctuations $\delta\phi /\phi_0 \sim \mathcal{O}(1)$.
\end{enumerate}
%%
% are satisfied with this potential.
Note that these conditions are not sufficient but just necessary.
To confirm the oscillon formation, 
we should perform classical lattice simulation.

%%%%%%%%%%%%%%%%%%%%%%%%%%%%%%%%%%%%%%%%%%%%%%
\subsection{Oscillon solution}
\label{sec:solution}

First, let us show that the oscillon configuration can be realized in the potential Eq.~(\ref{eq:potential}). 

Oscillon is the non-topological pseudo soliton 
approximately conserving the adiabatic invariant $I$~\cite{Kasuya:2002zs} defined as
\begin{equation}
    I \equiv \frac{1}{\omega} \int d^{3}x \overline{\dot{\phi}^{2}},
\end{equation}
where the overline represents the time average over periodic motion of $\phi$
and $\omega$ does its oscillation frequency almost the same as the mass of $\phi$.

The oscillon configuration is obtained by minimizing the time-averaged energy $\overline{E}$
for a given $I$.
With the use of the Lagrangian multiplier method,
\begin{align}
    E_{\lambda}
    &= \overline{E} 
    + \lambda \left( I - \frac{1}{\omega} \int dx^{3} \overline{\dot{\phi}^2} \right), \\
    &= \int d^{3}x \left[ \left(1-\frac{2\lambda}{\omega}\right)\frac{1}{2}\overline{\dot{\phi}^2}
    + \frac{1}{2} \overline{(\nabla\phi)^2} + \overline{V(\phi)} \right] + \lambda I.
    \label{eq:LM-energy}
\end{align}
Assuming that the periodic motion of $\phi$ is written as 
$\phi(x) \simeq \Phi(\bm{x}) \cos \omega t$ 
and defining $V(\Phi)\equiv2\overline{V(\phi)}$, Eq.~(\ref{eq:LM-energy}) becomes
\begin{align}
    E_{\omega} 
    &= \frac{1}{2} \int d^{3}x 
    \left[\omega \left(\omega-2\lambda\right) \frac{1}{2}\Phi^{2}
    +\frac{1}{2} (\nabla\Phi)^{2} + V(\Phi) \right] + \lambda I, \\
    & =\frac{1}{2} \int d^{3}x
    \left[\frac{1}{2}(\nabla\Phi)^{2}+V(\Phi)-\frac{1}{2}\omega^2\Phi^{2} \right]
    + \omega I.
\end{align}
where we use the relation $\lambda = \omega$~\cite{Ibe:2019vyo} in the last line. 
Because the lowest energy configuration is realized in the spherically symmetric configuration, 
we set $\Phi(\bm{x})=\Phi(r)$ and impose the boundary condition as
\begin{equation}
\left.\Phi\right|_{r\rightarrow\infty}=0,\ \ \ \left.\frac{d\Phi}{dr}\right|_{r\rightarrow0}=0.\label{eq:boundary condition}
\end{equation}
Under this condition, 
by differentiating $E_{\omega}$ by $\Phi$ to get the extremum we obtain 
\begin{equation}
    \frac{d^{2}\Phi}{dr^{2}}
    + \frac{2}{r} \frac{d\Phi}{dr}
    + \frac{d}{d\Phi} \left( \frac{1}{2}\omega^2\Phi^{2}-V(\Phi) \right)
    = 0.
    \label{eq:oscillon-eq}
\end{equation}
This equation is considered as the equations of motion of $\Phi$
moving in the potential $\omega^2\Phi^{2}/2-V(\Phi)$ with friction.
Thus the condition for existing the solution satisfying the boundary
condition Eq.~(\ref{eq:boundary condition}) is
\begin{equation}
    \min \left[ \frac{V(\Phi)}{\Phi^{2}}\right] < \omega^2 < m^2.
    \label{eq:oscillon-condition}
\end{equation}
Because the ULAP potential Eq.~(\ref{eq:potential}) of $p > -1$ satisfies Eq.~(\ref{eq:oscillon-condition}),
the oscillon configuration is realized when the ULAP fluctuations are large enough.

%%%%%%%%%%%%%%%%%%%%%%%%%%%%%%%%%%%%%%%%%%%%%%
\subsection{Instability growth}
\label{sec:Floquet}

Not only the existence of the oscillon solution but also the large spatial inhomogeneity is necessary for the oscillon formation.
In the ULAP case, however, the field value is almost uniform at the beginning of the oscillation
because the homogeneous initial value is set by inflation.
Therefore, we discuss the growth of the small initial fluctuations by Floquet analysis in this subsection.

For analytical purposes, 
we expand the potential Eq.~(\ref{eq:potential}) up to a quartic order under $\phi < F$,
\begin{equation}
    V(\phi) \simeq 
    \frac{m^2F^2}{2} \left( \frac{\phi^2}{F^2} 
    - \frac{p+1}{2} \frac{\phi^4}{F^4} \right).
\end{equation}

Assuming that the background is dominated by the harmonic oscillation $\phi_0 \simeq \Phi \cos (mt)$, 
the equation of motion of the $k$-mode fluctuations without cosmic expansion leads to the Mathieu equation,
\begin{align}
    \ddot{\delta \phi_k} + \left[k^2 + V''(\phi_0) \right] \delta \phi_k
    &= 0,\\
    \Leftrightarrow\ \ \ 
    \ddot{\delta \phi_k} 
    + \left[ k^2 + m^2 
    - \frac{3(p+1)}{2} m^2 \left(\frac{\Phi}{F}\right)^2 
    \left( 1 + \cos 2mt \right) 
    \right] \delta \phi_k
    &\simeq 0.
\end{align}
When $\Phi \lesssim F$, 
the narrow instability band is induced with the approximate growth rate (Floquet exponent)
\begin{equation}
    \frac{\mu_{\rm max}}{m} \simeq \frac{3(p+1)}{8} \left(\frac{\Phi}{F}\right)^2,
    \hspace{0.5cm} {\rm at} \hspace{0.4cm}
    \frac{k}{m} \simeq \sqrt{\frac{3(p+1)}{2}} \frac{\Phi}{F},
    \label{eq:Floquet_exponent}
\end{equation}
which shows that the larger $p$ leads to the stronger instability
\footnote{
    See Ref.~\cite{Olle:2019kbo} for Floquet exponent of other $p$.
}.
In an expanding universe, when $\mu_{\rm max}/H \gtrsim 1$,
the fluctuation growth beats the cosmic expansion
and it can lead to large enough fluctuations for the oscillon formation.

%%%%%%%%%%%%%%%%%%%%%%%%%%%%%%%%%%%%%%%%%%%%%%%%%%%%%%%%%%%%%
\section{Oscillon Formation}
\label{sec:formation}

In this section,
we verify the oscillon production by the classical lattice simulation
after a brief explanation of the simulation setup.
The oscillon formation in the potential Eq.~(\ref{eq:potential}) is confirmed in many papers in the inflationary context~\cite{Amin:2011hj, Lozanov:2017hjm, Kitajima:2018zco, Hong:2017ooe}.
Here, we test the formation in the radiation dominated universe.

%%%%%%%%%%%%%%%%%%%%%%%%%%%%%%%%%%%%%%%%%%%%%%
\subsection{Simulation setup}
\label{sec:setup}

In the simulation,
the units of the field, time, space, etc. are taken to be $F$ and $m^{-1}$, that is,
\begin{equation}
    \bar{\phi} \equiv \frac{\phi}{F},\ \
    \bar{\tau} \equiv m\tau,\ \
    \bar{x} \equiv mx,\ \ \dots\ {\rm etc}.
\end{equation}
where the overline denotes the dimensionless program variables and $\tau$ is the conformal time.

The equation of motion of $\bar{\phi}$ is represented by
\begin{equation}
    \bar{\phi}'' + 2 \frac{a'}{a}\bar{\phi}' 
    - \bar{\Delta} \bar{\phi} 
    + a^2\frac{\partial \overline{V}}{\partial \bar{\phi}} = 0,
    \hspace{0.6cm}
    \overline{V}
    = \frac{1}{2p} \left[ 1 - \left(1+\bar{\phi}^2\right)^{-p} \right].
    \label{eq:num_eom}
\end{equation}
where dash denotes the derivative of $\overline{\tau}$.
As the initial condition,
%%%%%%%%%%%%%%%%%%%%%%%%%%%%
% \vspace{-0.2cm}
\begin{itemize}
    \item Hubble parameter:
        ULAP starts to oscillate in the radiation dominated universe, so we take $H_i = 1/2t = m$.
    \item Scale factor:
        the initial scale factor is set to be unity $a_i = 1$.
        (in the radiation dominated universe $a = \bar{\tau}$).
    \item Field values: 
        the initial field value and its derivative are set as
        \begin{eqnarray}
            \bar{\phi}_i = \frac{2}{3}\pi (1 + \zeta),\ \ \bar{\phi}'_i = 0.
        \end{eqnarray}
        where $\zeta$ is the uniform random fluctuations of $\mathcal{O}(10^{-5})$
        \footnote{
            The initial field average $\langle \bar{\phi}_i \rangle = 2\pi/3$ is just the benchmark.
        }.
\end{itemize}
% \vspace{-0.2cm}
%%%%%%%%%%%%%%%%%%%%%%%%%%%%
The other simulation parameters are shown in Table~\ref{Ta:params}.

%%%%%%%%%%%%%%%%%%%%%%%%%%%%
\begin{table}[t]
    \centering
    \begin{tabular}{cc}
        \hline \hline
        $p$ & varying \tabularnewline
        Initial field value $\langle \bar{\phi}_i \rangle$ & $2\pi/3$ \tabularnewline
        Box size $L$  & $8$ \tabularnewline
        Grid size $N$ & $256^3$ \tabularnewline
        Final time    & $50$ \tabularnewline
        Time step     & $0.01$ \tabularnewline
        \hline \hline
    \end{tabular}
    \caption{
        Simulation parameters.
        $p$ is changed in every simulation to set the appropriate potential.
    }
    \label{Ta:params}
\end{table}
%%%%%%%%%%%%%%%%%%%%%%%%%%%%

We utilize our lattice simulation code used in Refs.~\cite{Ibe:2019vyo, Ibe:2019lzv},
in which the time evolution is calculated by the fourth-order symplectic integration scheme 
and the spatial derivatives are calculated by the fourth-order central difference scheme.
We impose the periodic boundary condition on the boundary.
% We have confirmed that the results do not depend on the box size $L$, grid size $N$, and the time step $\Delta t$.

%%%%%%%%%%%%%%%%%%%%%%%%%%%%%%%%%%%%%%%%%%%%%%
\subsection{Simulation results}
\label{results}

%%%%%%%%%%%%%%%%%%%%%%%%%%%%
\begin{figure}[t]
    \begin{minipage}{.48\linewidth}
        \begin{center}
            \includegraphics[width=\linewidth]{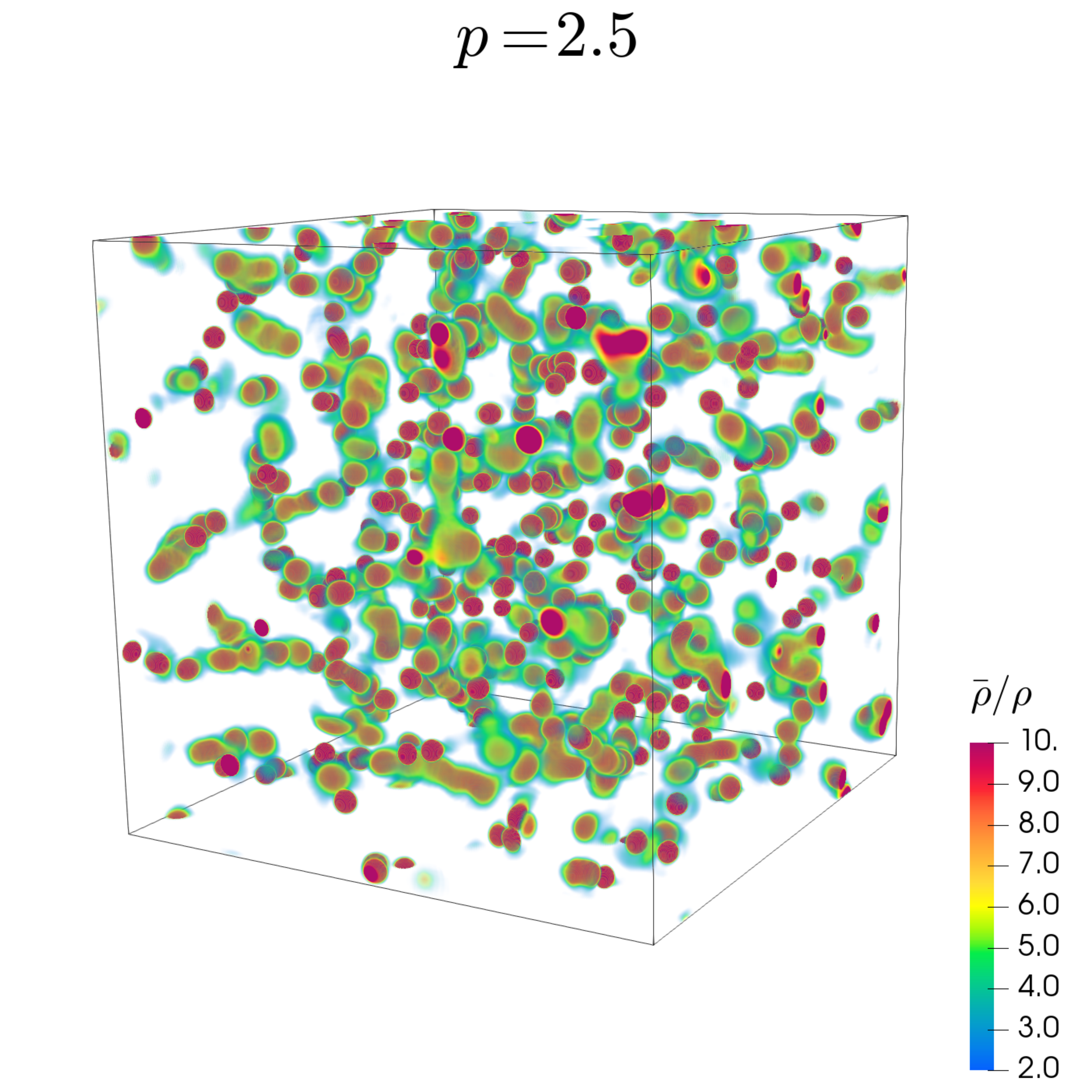}
        \end{center}
    \end{minipage}
    \begin{minipage}{.48\linewidth}
        \begin{center}
            \includegraphics[width=\linewidth]{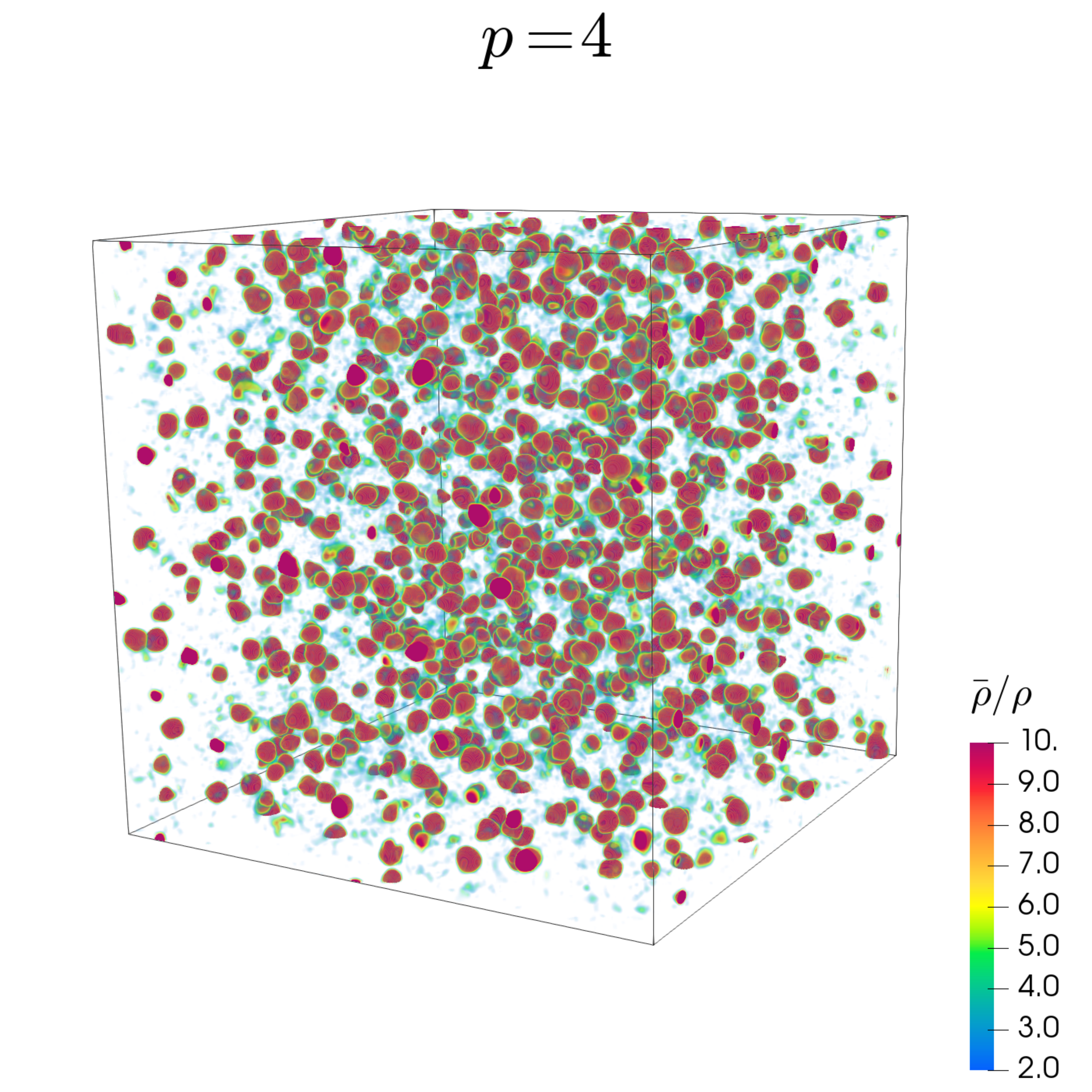}
        \end{center}
    \end{minipage}
    \caption{
        Simulation results. 
        We plot the ratio of the energy density $\rho$ to the averaged one $\bar{\rho}$ in the simulation box at $a = 30$.
        Left and right figures show the case of $p = 2.5$ and $4$ respectively.
        From these figures,
        We can find that a lot of oscillons are produced by self-resonance.
    }
    \label{fig:sim_results}
\end{figure}
%%%%%%%%%%%%%%%%%%%%%%%%%%%%

The simulation results are shown in Fig.~\ref{fig:sim_results} for $p = 2.5$ (left) and $p = 4$ (right).

In both cases, the high energy local objects are produced.
However, we have confirmed by simulations that the oscillon formation does not occur
when $p \lesssim 2$.
There are two main reasons for this. 
The first is that smaller $p$ leads to the weaker resonance as shown in Eq.~(\ref{eq:Floquet_exponent}).
The twice larger $\mu$ after few oscillations makes fluctuations $e^{10} \sim 10^4$  times larger.
The second is that 
the effective mass at the beginning of the oscillation $m_{\rm eff}^2 = V''(\phi_i)$ becomes smaller in larger $p$.
This causes the later onset of the oscillation,
which leads to the increase of the number of oscillations in each Hubble time 
and instability enhancement.

We also comment on the initial field value.
In this simulation, we set $\langle \bar{\phi}_i \rangle = 2\pi/3$ as the benchmark of the $\mathcal{O}(1)$ initial field value.
However, if we assume the highly fine-tuned initial condition as 
$\langle \bar{\phi}_i \rangle = \mathcal{O}(10^{-1})$ or so,
fluctuations may not be grown enough because of the opposite reasons to the above two.
On the other hand, 
if we assume the larger initial value,
even smaller $p$ may cause the oscillon formation.

Although we focused only on the potential Eq.~(\ref{eq:potential}) in this paper,
other ULAP potentials satisfying the condition Eq.~(\ref{eq:oscillon-condition}) also have the possibility of the oscillon formation.

%%%%%%%%%%%%%%%%%%%%%%%%%%%%%%%%%%%%%%%%%%%%%%%%%%%%%%%%%%%%%
\section{Oscillon Lifetime}
\label{sec:lifetime}

In this section, we briefly estimate the oscillon lifetime based on Ref.~\cite{Ibe:2019vyo}
and discuss the effect of the resultant oscillon.
% For more details of the lifetime derivation, see Ref.~\cite{Ibe:2019vyo}.

In Ref.~\cite{Ibe:2019vyo},
we derive the method to calculate the classical decay rate of oscillon 
by solving the equation of motion of fluctuations around the theoretical oscillon profile.
Let us decompose a scalar field $\phi(x)$ into the oscillon profile and fluctuations as
\begin{equation}
    \phi(x) = \psi(r) \cos \omega t + \xi(x).
\end{equation}
where $\psi(r)$ obeys the oscillon profile equation Eq.~(\ref{eq:oscillon-eq})
\footnote{
    We impose the spherical symmetry on the system 
    because we focus only on the lowest energy state.
}. 
With the use of the equation of motion of $\phi$ and Eq.~(\ref{eq:oscillon-eq}),
the equation of motion of $\xi$ is solved and
the energy loss rate of oscillon 
\begin{equation}
    \frac{dE}{dt} = 4\pi r^2 T^{0r},
\end{equation}
where $T_{0r} = \partial_0 \xi \partial_r \xi$ denotes the Poynting vector
can be calculated.

%%%%%%%%%%%%%%%%%%%%%%%%%%%%
\begin{figure}[t]
    \begin{minipage}{.45\linewidth}
        \begin{center}
            \includegraphics[width=\linewidth]{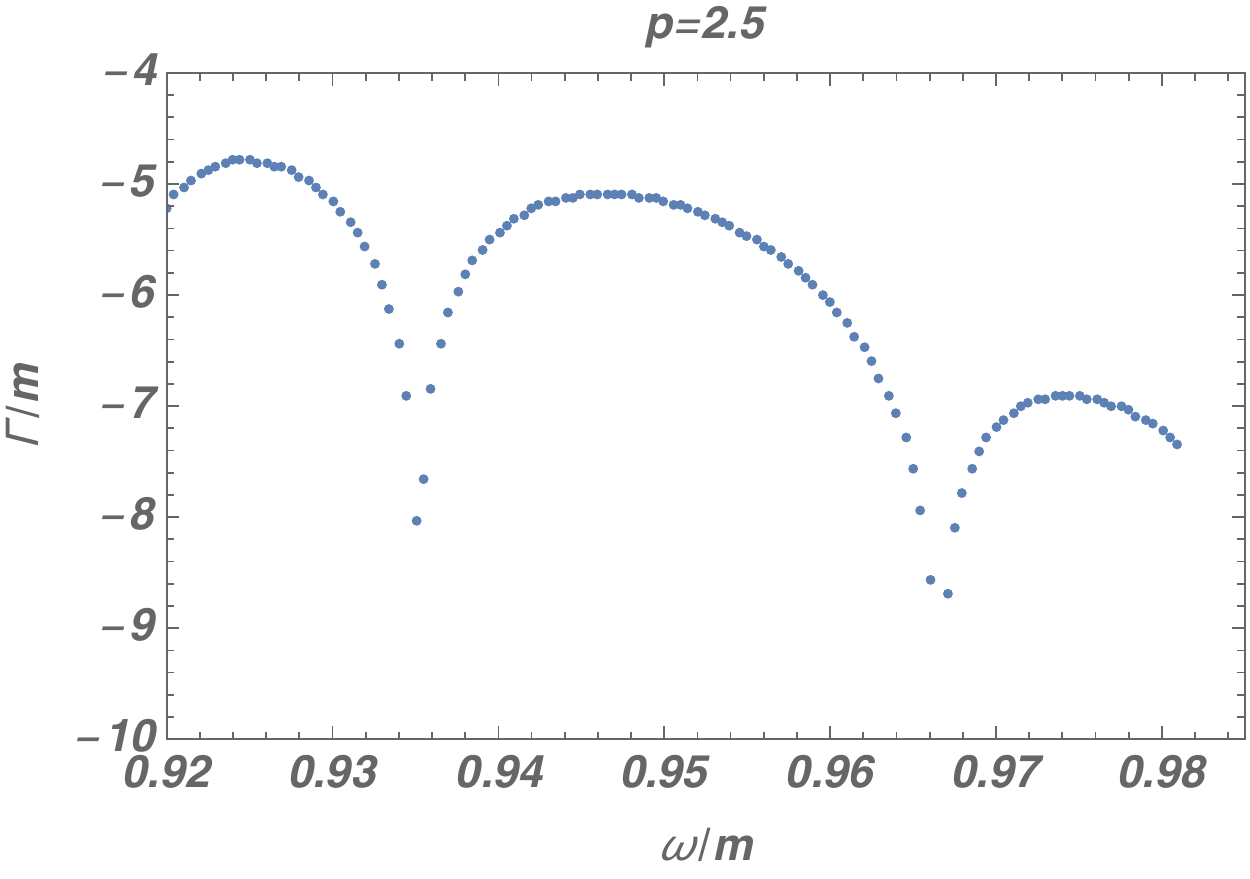}
        \end{center}
    \end{minipage}
    \hspace{.3cm}
    \begin{minipage}{.45\linewidth}
        \begin{center}
            \includegraphics[width=\linewidth]{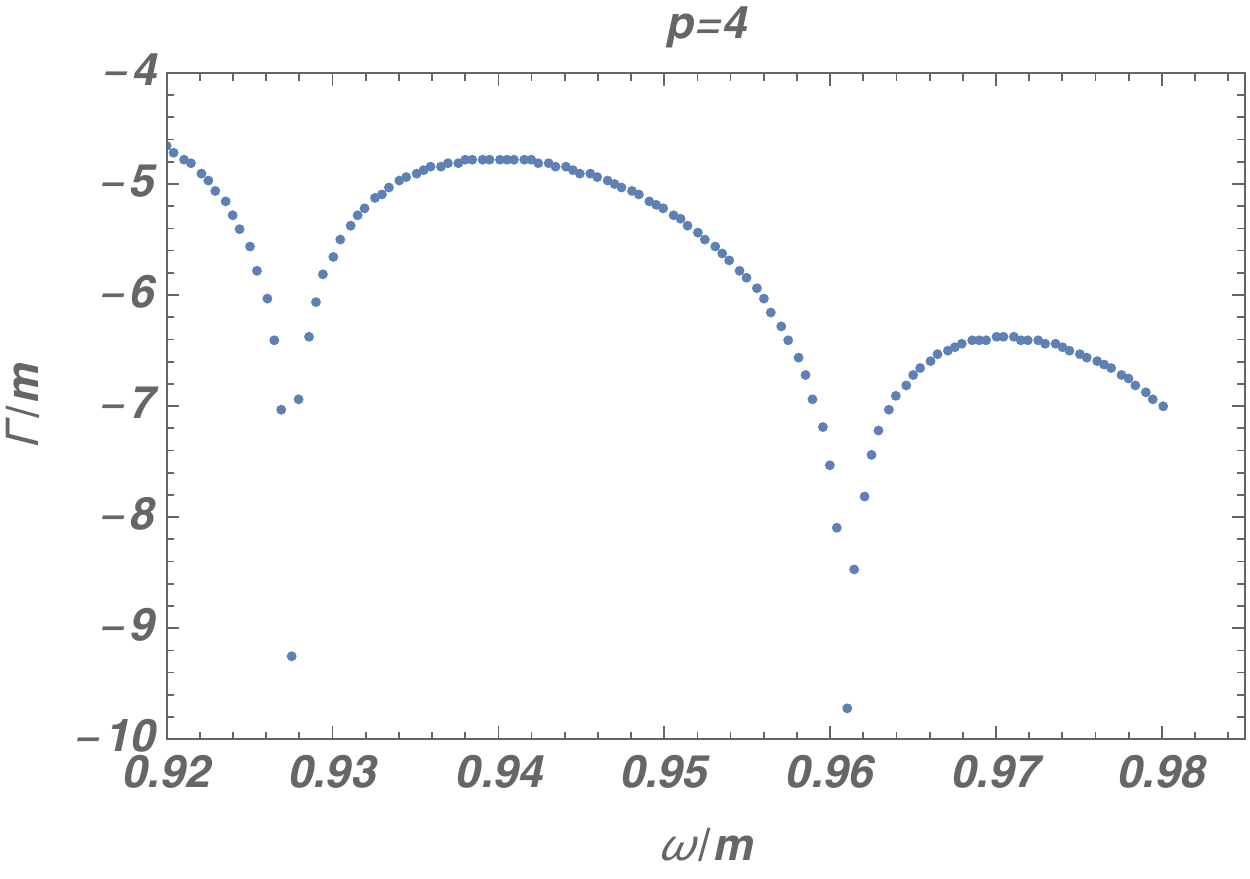}
        \end{center}
    \end{minipage}
    \caption{
        The decay rate of oscillon.
        The left and right figures show the case of $p = 2.5, 4$ respectively.
        For $\omega \gtrsim 0.98$ the oscillon does not satisfy the stability condition $\frac{\partial \omega}{\partial I} > 0$ \cite{Mukaida:2016hwd}. 
        Larger $\omega$ means the smaller oscillon by definition,
        so the minimum lifetime is $\mathcal{O}(10^7)m$.
        In addition, if larger osillons (e.g., $\omega \lesssim 0.96$) are formed,
        the lifetime may be much longer 
        because the decay rate becomes much smaller there.
    }
    \label{fig:lifetime}
\end{figure}
%%%%%%%%%%%%%%%%%%%%%%%%%%%%

The derived decay rate $\Gamma = |\dot{E}|/E$ is shown in Fig.~\ref{fig:lifetime}.
Since the larger $\omega$ corresponds to the smaller oscillon (see Sec.~\ref{sec:solution} for the definition of $\omega$), the decay rate of a large oscillon evolves following the curve in Fig.~\ref{fig:lifetime} from left to right.
From Fig.~\ref{fig:lifetime} it is found that the produced oscillon is stable at least for $\mathcal{O}(10^7)m$. 
For the typical ULAP mass $m \sim 10^{-22}\ {\rm eV}$,
the minimum lifetime is about $10^8$ years.
%These objects may result in completely new hybrid objects 
%which supported by both self and gravity interactions~\cite{Ikeda:2017qev, Ikeda:2018dav}.
Thus, they may affect the cosmic evolution, particularly structure formation.
%but oscillon evolution under gravity is unclear and it remains for our future work.

In addition, 
the lifetime of oscillon may become longer 
because its lifetime depends on its size.
As shown in the figures, the decay rate has a lot of poles
where the decay rate is extremely small.
At these points, oscillon hardly evolves to the smaller one,
which results in a longer lifetime ($\gtrsim 10^{10}$ years).
Therefore, if the large oscillons are formed, 
%and it is not disrupted by gravity,
ULAP oscillons may still live in the present universe.

%%%%%%%%%%%%%%%%%%%%%%%%%%%%%%%%%%%%%%%%%%%%%%%%%%%%%%%%%%%%%
\section{Conclusions}
\label{sec:conclusion}

In this paper, we have examined the oscillon formation in a ULAP potential, Eq.~(\ref{eq:potential}).
We have shown that 
oscillons are really produced when the potential index is $p > 2$ with $\langle \phi_i \rangle/F = 2\pi/3$.
Their lifetime is at least $\mathcal{O}(10^7)m$, which equals to $10^8$ years with $m \sim 10^{-22}\ {\rm eV}$.
Because such long lifetime objects survive until the structure formation,
they may affect dynamical history of the universe.
%produce completely new objects by self-interaction and gravity.

Moreover, the lifetime could be much longer if large oscillons are formed
because the decay rate of the oscillon extremely decreases in the specific profiles.
% The oscillon existence in the present universe may affect current ULAP observations.
In either case,
a deep understanding of the evolution of ULAP oscillons is indispensable to consider effects on cosmology.
Detailed study should be discussed in the future work.

%%%%%%%%%%%%%%%%%%%%%%%%%%%%%%%%%%%%%%%%%%%%%%%%%%%%%%%%%%%%%
\begin{acknowledgments}

This work was supported by JSPS KAKENHI Grant Nos. 17H01131 (M.K.) and 17K05434 (M.K.), MEXT KAKENHI Grant Nos. 15H05889 (M.K.), World Premier International Research Center Initiative (WPI Initiative), MEXT, Japan, and JSPS Research Fellowships for Young Scientists Grant No. 19J12936 (E.S.).

\end{acknowledgments}
%%%%%%%%%%%%%%%%%%%%%%%%%%%%%%%%%%%%%%%%%%%%%%%%%%%%%%%%%%%%%

\bibliographystyle{JHEP}
\bibliography{refs}
\end{document}